\def\be{\begin{eqnarray}}
\def\en{\end{eqnarray}}

\def\la{\langle}
\def\ra{\rangle}
\def\ov{\overline}
\def\B{{\cal B}}
\def\P{{\cal P}}

\documentclass{ws-ijmpa}

\begin{document}

\markboth{Hai-Yang Cheng} {LIGHT-FRONT APPROACH FOR STRONG AND
WEAK DECAYS OF PENTAQUARKS}

\catchline{}{}{}{}{}
%
%%%%%%%%%%%%%%%%%%%%%%%%%%%%%%%%%%%%%%%%%%%%%%%%%%%%%%%%%%%%%%%%%%%%

\title{LIGHT-FRONT APPROACH FOR STRONG AND WEAK DECAYS OF PENTAQUARKS\\
}

\author{\footnotesize Hai-Yang Cheng}

\address{Institute of Physics, Academia Sinica\\
Taipei, Taiwan 115, ROC}

\maketitle

%\pub{Received (Day Month Year)}{Revised (Day Month Year)}

\begin{abstract}

Strong and weak decays of pentaquarks are studied in the framework
of the light-front approach.

\end{abstract}

\section{Introduction}
The discovery of an exotic $\Theta^+$ pentaquark by LEPS at
SPring-8,\cite{LEPS} subsequently confirmed by many other groups,
marked a new era for testing our understanding of the hadron
spectroscopy and promoted a re-examination of the QCD implications
for exotic hadrons. The mass of the $\Theta^+$ is of order 1535
MeV and its width is less than 10 MeV from direct observations and
can be as narrow as 1 MeV or even lower.\cite{width} Many null
results for the pentaquark search mostly from high energy
experiments have also been reported. Therefore, if the $\Theta^+$
pentaquark is real, it must be established beyond any doubt.

If the $\Theta^+$ pentaquark exists, its minimum quark content is
$uudd\bar s$. To understand those experimental measurements with
positive results, we are facing three puzzles: (i) The doubly
charged partner of $\Theta^+$, namely, $\Theta^{++}$ with the
quark content $uuud\bar s$, should be easily detected via the
decay $\Theta^{++}\to K^+p$. The puzzle is why is it not seen so
far while stringent limits have been set ? (ii) The naive
constituent quark model in which quarks are uncorrelated implies a
$\Theta^+$ pentaquark mass of order 1900 MeV. Why is it so
anomalously light ? (iii) The fall-apart strong decay $\Theta^+\to
KN$ is OZI super-allowed. Hence, if quarks in $\Theta^+$ are
uncorrelated, the width will be of order several hundred MeV. Why
is the observed $\Theta^+$ width so narrow ? These puzzles hint at
a possible correlation among various quarks; two or three quarks
could form a cluster. A popular correlated quark model has been
advocated by Jaffe and Wilczek \cite{JW} in which the $\Theta^+$
is a bound state of an $\bar s$ quark with two $(ud)$ diquarks.
The diquark is a highly correlated spin-zero object and is in a
flavor anti-triplet and color anti-triplet state. Bose statistics
of the scalar diquarks requires that the diquark pairs be in an
orbital $P$ wave state. The parity of $\Theta^+$ is predicted to
be positive owing to the diquark correlation.

\section {Strong decays of pentaquarks}

Consider the strong decay $\Theta^+\to K^+n$. It can proceed
through two processes: fall-apart decay with quark annihilation
and the kaon emission process. The latter is expected to be
severely suppressed (except in the infinite momentum frame where
quark annihilation and annihilation are prohibited) as it involves
the transition between $\Theta^+$ and the 5-quark component of the
nucleon. The fall-apart process is quite unique to the pentaquark
and it cannot occur in the ordinary baryon decays. As mentioned
before, if quarks are uncorrelated, the fall-apart mechanism will
yield a width of several hundred MeV. Therefore, quark correlation
is needed to suppress the OZI supper-allowed process. In the
Jaffe-Wilczek model, this suppression can be understood as
follows. First, one has to break the $ud$ diquark pair so that one
of the light quarks will combine with the $\bar s$ quark to form a
kaon. Second, one has to overcome the $P$-wave potential barrier
so that the rest three quarks form a $s$-wave nucleon. Hence, the
$\Theta^+$ decay width is narrow due to the breaking of diquarks
and the transformation of the $p$-wave into $s$-wave
configurations.

In a typical pentaquark decay to a meson and a baryon, the
anti-quark is common to both pentaquark and the final state meson.
To the leading order of the spectator approximation, the
anti-quark can be considered as a spectator in the decay process
depicted in Fig.~1. In this picture, there is a  $\phi\phi\to
{\cal B} q$ subprocess with $\phi\phi$ being a diquark pair and
${\cal B}$ a baryon. We use the effective Hamiltonian \cite{CC}
 \be \label{eq:Heff}
 H_{\rm eff}&=&\frac{g_1}{M}
 ~\ov {\cal B} \gamma_5 q^c
 \phi\phi
 +\frac{g_{2}}{M^2}
 ~\ov {\cal B} i\gamma^\mu\gamma_5(q^c)
 \phi\partial_\mu\phi
 \en
to model the $\phi\phi\to {\cal B} q$ subprocess, where $M={\cal
O}(m_\phi, m_{\cal B})$ is a characteristic scale of the system.
It turns out that only the $g_2$ term contributes to even-parity
pentaquark decays. Although we do not know how to calculate
$g_2/M^2$ from first-principles calculations, we can determine it
from the $\Theta$ width, and use it to predict the decay widths of
other light and heavy pentaquarks.

\begin{figure}[t]
\centerline{
            {\epsfxsize1.2 in \epsffile{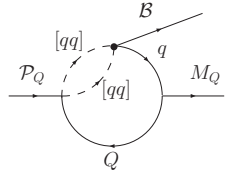}}}
\caption{Feynman diagram for a typical $\P_Q\to M_{Q}{\cal B}$
transition with ${\cal B}$ being an octet baryon, where the
spin-zero diquarks $([qq]=[ud],[us],[ds])$ are denoted by dashed
lines and the corresponding operator ${\cal O}_{\rm eff}$.}
\label{fig:penta} %%
\end{figure}

For the strong decay $\P\to M\B$, we have calculated the matrix
element $A(\P\to M\B)=(g_2/M^2)\bar u\la
M|i\gamma^\mu\gamma_5q^c\phi\partial_\mu\phi|\P\ra$. The matrix
elements can be expressed in terms of form factors. We first
computed the form factors in the spacelike region and then
extrapolated them to the timelike region. Taking 1 MeV as a
benchmark for the $\Theta$ width, it is found that $g_2/M^2\approx
10.2\,{\rm GeV}^{-2}$. We have estimated the strong decays
$\Xi^{--}_{3/2}\to\Xi^-\pi^-,\Sigma^-K^-$, $\Theta^0_c\to pD^-$,
$\Sigma_{5c}\to p D^-_s, pD^{*-}_{s0}$ and $\Xi^0_{5c}\to \Sigma^+
D^-_s,pD^{*-}_{s0}$ by normalizing to the $\Theta^+$ width. We
found that $\Xi^{--}_{3/2}\to\Xi^-\pi^-,\Sigma^-K^-$,
$\Sigma_{5c}\to p D^-_s, pD^{*-}_{s0}$ and $\Theta^0_c\to
pD^-$decay rates are of the order of a few MeV, while
$\Xi^0_{5c}\to \Sigma^+ D^-_s, \Sigma^+D^{*-}_{s0}$ decay rates
are of order tens of MeV.\cite{CC} If we take
$m_{\Xi^{--}_{3/2}}=1862\pm 2$~MeV as observed by NA49,\cite{NA49}
we will have
$\Gamma(\Xi^{--}_{3/2}\to\Xi^-\pi^-)/\Gamma(\Theta^+\to
pK^0)\simeq 2.2$ which is consistent with the observed width of
$\Gamma(\Xi^{--}_{3/2})\leq 18$~MeV.\cite{NA49}

\section {Weak decays of heavy pentaquarks}
If the light $\Theta^+$ is real, it is expected the existence of
the stable heavy pentaquark $\Theta_Q$ by replacing $\bar s$ by
$\bar Q$. Whether the heavy pentaquark lies above or below the
strong decay threshold is rather controversial. A narrow resonance
in $D^*p$ mass distribution with mass $3099\pm3\pm5$ MeV and width
$12\pm3$ has been reported by H1,\cite{H1} but it has not been
confirmed by many other groups. If the heavy pentaquarks lie below
the strong decay threshold, they can be searched for only through
weak or electromagentic decays.

We apply the relativistic light-front approach to calculate the
pentaquark to pentaquark  transition form factors.\cite{CCH} In
the heavy quark limit, heavy-to-heavy pentaquark transition form
factors can be expressed in terms of three Isgur-Wise functions:
two of them are found to be normalized to unity at zero recoil,
while the third one is equal to 1/2 at the maximum momentum
transfer, in accordance with the prediction of the large-$N_c$
approach \cite{Chow} or the quark model.\cite{Chengbottom}
Therefore, the light-front model calculations are consistent with
the requirement of heavy quark symmetry.

We have calculated form factors and Isgur-Wise functions in
\cite{CCH}. Decay rates of the weak decays
$\Theta_b^+\to\Theta_c^0 \pi^+(\rho^+)$, $\Theta_c^0\to\Theta^+
\pi^-(\rho^-)$, $\Sigma'^+_{5b}\to \Sigma'^0_{5c}\pi^+(\rho^+)$
and $\Sigma'^0_{5c}\to N^+_8\pi^-(\rho^-)$ with $\Theta_Q$,
$\Sigma'_{5Q}$ and $N_8$ being the heavy anti-sextet, heavy
triplet and light octet pentaquarks, respectively, are obtained.
For weakly decaying $\Theta_b^+$ and $\Theta_c^0$, the branching
ratios of $\Theta_b^+\to\Theta_c^0\pi^+$,
$\Theta_c^0\to\Theta^+\pi^-$ are estimated to be at the level of
$10^{-3}$ and a few percents, respectively.

\section{Acknowledgments}

I am grateful to Chun-Khiang Chua and Chien-Wen Hwang for fruitful
collaboration.

\end{document}